# Synergistically creating sulfur vacancies in semimetal-supported amorphous MoS₂ for efficient hydrogen evolution

Guowei Li,[a*] Chenguang Fu,[a] Jiquan Wu,[b] Jiancun Rao,[c] Sz-Chian Liou,[c] Xijin Xu,[d] Baiqi Shao,[e] Kai Liu,[e] Enke Liu,[f] Nitesh Kumar,[a] Xianjie Liu,[b] Mats Fahlman,[b] Johannes Gooth,[a] Gudrun Auffermann,[a] Yan Sun,[a] and Claudia Felser,[a] and Baomin Zhang [a*]

[a] *Max Planck Institute for Chemical Physics of Solids, 01187 Dresden, Germany*

[b] *Department of Physics, Chemistry and Biology (IFM), Linköping University, 58183 Linköping, Sweden*

[c] *AIM Lab, Maryland NanoCenter, University of Maryland, MD 20742, USA*

[d] *Changchun Institute of Applied Chemistry, Chinese Academy of Sciences, 130022, Changchun, China.*

[e] *Institute of Physics, Chinese Academy of Sciences, 100190 Beijing, China*

[f] *School of Physics and Technology, University of Jinan, 250022 Jinan, China*

**Abstract:** The presence of elemental vacancies in materials are inevitable according to statistical thermodynamics, which will decide the chemical and physical properties of the investigated system. However, the controlled manipulation of vacancies for specific applications is a challenge. Here we report a facile method for creating large concentrations of S vacancies in the inert basal plane of MoS₂ supported on semimetal CoMoP₂. With a small applied potential, S atoms can be removed in the form of H₂S due to the optimized free energy of formation. The existence of vacancies favors electron injection from the electrode to the active site by decreasing the contact resistance. As a consequence, the activity is increased by 221 % with the vacancy-rich MoS₂ as electrocatalyst for hydrogen evolution reaction (HER). A small overpotential of 75 mV is needed to deliver a current density of 10 mA cm$^{-2}$, which is considered among the best values achieved for MoS₂. It is envisaged that this work may provide a new strategy for utilizing the semimetal phase for structuring MoS₂ into a multi-functional material.

## 1. Introduction

The elemental vacancies in materials, which are generally called defects, are considered to be perfect and powerful tools for designing various functional materials.[1-4] The vacancies can be viewed as virtual atoms that have an empty electronic state at the vacuum level, and thus the charge imbalances will lead to the redistribution of charges and consequently induce defect levels in the band gap, resulting in a richness of phenomena such as band gap narrowing, band bending, and symmetry distortion.[5] As a typical two-dimensional (2D) material, $MoS_2$ provides an ideal prototype to explore the interaction between vacancies and functions, especially in the field of electrochemical water splitting.[6-11] It has been proven that the edge sites are catalytically active, while the basal plane is pretty inert.[12] Recent work demonstrated that sulfur vacancies could serve as another important catalytically active site for HER due to the favorable adsorption free energy.[13, 14] This is supported by the theoretical calculations of the formation energy of S vacancies, which is much lower than that of Mo vacancies. The major result is the emergence of impurity states in the bandgap that allow favorable hydrogen adsorption.[15-17]

However, the controlled fabrication of S vacancies in the inert plane of $MoS_2$ is a formidable challenge and generally needs critical conditions such as electron/argon irradiation, hydrogen plasma treatment, or high-temperature annealing.[18, 19] A recent work reported the possibility of creating S vacancies in the electrochemical process but need a large accessible applied potential. This is understandable because the removal of S by $H_2$ to release $H_2S$ is very endothermic and kinetically difficult to achieve.[20, 21] Recent work found that the robust electronic states on topological insulators could enhance the adsorption of various molecular species when covered by selected catalytic metal layers.[22-24] Informed by the deposition of oxide thin films on the substrate, where oxygen vacancies can be formed due to the strong affinity between oxygen and the element from the substrate,[25] it is very interesting to explore the synergistic effect on the formation of S vacancies in the semimetal phase – supported $MoS_2$.

As a recently defined semimetal, $CoMoP_2$ is selected as the supporting substrate because of the good conductivity and similar hexagonal structure as $MoS_2$.[26] Density functional theory (DFT) calculation predicted that the S atoms in the basal plane could be activated and removed by bonding with hydrogen. The formation energy barrier for $H_2S$ was significantly decreased in comparison with the pure $MoS_2$ phase without support. S vacancies with a concentration of 13 % could be created in the electrochemical process. Consequently, the catalytic activity was increased by 221 % at a given overpotential (-0.25 V *vs* RHE) and show high stability in a wide potential window. This opens new pathways for the ready creation of vacancies in 2D materials for various applications.

## 2. Experimental

*2.1. Synthesis of the catalysts*

The topological semimetal $CoMoP_2$ was synthesized by a solid state reaction. In a typical synthesis, 0.25 mmol $NH_4Mo_7O_{24}$ $4H_2O$, 1.75 mmol $Co(NO_3)_2 \cdot 6H_2O$, 4 mmol $NH_4Mo_7O_{24} 4H_2O$, 4 mmol $(NH_4)_2HPO_2$, 4 mmol citric acid, and 2.5 g urea were dissolved in distilled water. Then the solution was dried at 80°C and then heated at 500 °C for 4 h in the air. After this, it was moved into a tube furnace and heated at 850 °C for 2 h with a heating rate of 1 °C / min. Hydrogen flow was used in this process.

For the synthesis of $MoS_2/CoMoP_2$ heterostructure (CMPS), 50 mg of the as-synthesized powder, 50 mg of $(NH_4)_2MoS_4$, and a Ni foam were placed in a 50 mL autoclave filled with 20 mL of N, N-Dimethylformamide. 0.1 mL of $N_2H_4$ $H_2O$ was added drop by drop. The mixed solution was transferred into an oven and heated at 180 °C for 24 h. The produced product was washed with water first and then with ethanol and dried at 60 °C for the following characterization.

*2.2. Characteristic techniques and electrochemical activation strategy*

Powder X-ray diffraction (XRD) data were recorded with a Bruker D8 Advance diffractometer equipped with a Cu Kα source (λ = 0.15406 nm). The morphologies and structures of the products were characterized by TITAN 80/300 electron microscope. XPS investigation was carried on a UHV surface analysis system equipped with a Scienta-200 hemispherical analyzer. The base pressure of a sample analysis chamber is $2 \times 10^{-10}$ mbar.

Electrochemical activation and performance assessment were performed on the Autolab PGSTAT302N with an impedance module electrochemistry workstation. A conventional three-electrode cell configuration was employed. The Ni foam with pristine CMPS sample was used directly as a working electrode. A Ag/AgCl (3 M KCl) electrode was used as the reference electrode, and a graphite rod was used as the counter electrode. 1 M KOH was used as the electrolyte. Linear sweep voltammetry was recorded at a scan rate of 1 mV/s. All the polarization curves were iR-corrected. The activation process was performed in potentiostatic mode at −0.126 V vs RHE for 40 h.

All the potentials reported in this work were converted to a reversible hydrogen electrode according to $E$ (versus RHE) = $E$ (versus Ag/AgCl) + (0.207 + 0.059 pH) V.

*2.3. Theoretical calculations*

The crystal structure of $CoMoP_2$ is fully relaxed (including the shape and the volume of the unit cell, and the internal positional parameters), and the optimized lattice parameters agree well with experimental values within a discrepancy of 0.5%. The optimized lattice parameters of $CoMoP_2$ in the *a-b* plane were used to construct the basic unit of the supercell in *a-b* plane. Taking account of the necessity of creating S vacancies in supercells and the acceptable computational efforts of DFT methods when treating vacancies in solids, in *a-b* plane, the dimension of the supercell is chosen to be 4*4 basic units of $CoMoP_2$. The heterostructure is constructed by adding one monolayer of $2H-MoS_2$ on top of the Co-terminated $CoMoP_2$ (001) surface. All calculations on supercells in this work are based on the slab model, and the thickness of the vacuum region is chosen to be as large as 12 Å to reduce the mirror image effect to a negligible value. In this work, we keep the shape and the volume of supercells fixed, optimize the internal positional parameters of supercells only and keep the Co atomic layer at the bottom of each supercell fixed mimicking the $CoMoP_2$ substrate.

## 3. Results and discussions

*3.1. Theoretical prediction of S vacancy formation*

To explore why and how S vacancies are formed, we first performed DFT calculations of the surface free energies on three different supercells: (a) In the case with one S atom adsorbed on the S vacancy (denoted as $^\square S$, which can also be called pristine $MoS_2/CoMoP_2$ heterostructure (CMPS), as shown in Figure S1a); (b) In the case with one S vacancy (denoted as $Mo^\square S_x/CoMoP_2$, as shown in Figure S1b), and (c) For the case with one H adsorbed on the S vacancy (denoted as $Mo^\square S_x(H)/CoMoP_2$, as shown in Figure S1c). As shown in Figure 1b, the surface energy of $Mo^\square S_x/CoMoP_2$ is larger than that of pristine structure, until the potential of as low as -1.26 V *vs* RHE. This is inconsistent with previous research, which indicated that despite perfect 2D materials such as $MoS_2$, is predicted to be unstable upon thermal fluctuation,[27] the creation of vacancies in the basal plane is reasonably difficult.[28] However, once created, they can be occupied immediately by hydrogen atoms under the electrochemical conditions in a broad potential window below 0.42 V *vs* RHE as illustrated by the arrow in Figure

1b. This value is much higher than that of pure $MoS_2$ phase, with a value of -0.26 V *vs* RHE.[20] This readily explains the instability of S vacancies because they can be easily passivated by adsorbates.[29] The vacancies created while supported on semimetal $CoMoP_2$ are more stable than the pure $MoS_2$ phase. We then calculated the Gibbs free energy for $H_2S$, which is the desulfurization process of making S vacancies. This is a two proton-electron transfer process, as shown in Figure 1 a and b. The transfer of the first electron and proton needs the energy of 1.43 eV, while the second protonation to form adsorbed $H_2S$ is uphill by about 1.51 eV from the first step. However, the whole reaction becomes exergonic with an absolute applied potential in the electrochemical environment. The ΔG for the formation of gaseous $H_2S$ was decreased to only 0.08 eV for the $MoS_2/CoMoP_2$ heterostructure, which is much lower than that for pure $MoS_2$ with a value of -0.83 eV. This means that the S can be removed more efficiently as $H_2S$ gas and serves as activity centers in the following HER process.

*3.2. Catalyst structure and composition*

Based on the calculations, we constructed the hybrid structure by a two-step process as illustrated in Figure 1d. $CoMoP_2$ nanoparticles were synthesized first by the in-situ reduction of Co-Mo complex precursor at high temperature under $H_2$ flow. Then, the pristine $MoS_2$ was directly grown on a Ni foam in the solution containing $CoMoP_2$ nanoparticles and ammonium thiomolybdate (($NH_4)_2MoS_4$). As revealed in the X-ray diffraction pattern (Figure 2a), the main peaks can be assigned to the hexagonal phase of $CoMoP_2$ with a space group $P6_3/mmc$, No.194 ($a$ = 0.333 nm, $c$ = 1.122 nm). The absence of diffraction peaks of $MoS_2$ proves that the obtained pristine phase is amorphous. Figure 2b shows the Scanning electron microscope (SEM) image of the final structure, with $MoS_2/CoMoP_2$ particles grown on the Ni foam. The corresponding energy-dispersive X-ray spectroscopy (EDS) analysis confirmed the particle is composed of the elements Mo, S, Co, and P (Figure 2b), which were distributed homogeneously according to elemental mapping (Figure S2). High-resolution transmission electron microscopy (HRTEM) further confirmed the crystallinity and composition of the as-synthesized hybrid structure. Several nanoparticles are covered by a thick layer of the amorphous shell (Figure 2c), but one can observe some tiny layered structures at the edge with a lattice fringe spacing of ~ 0.6 nm, corresponding to the (002) facet of bulk $2H-MoS_2$ (Figure 2d), indicating the formation of a few crystalline nuclei from the amorphous matrix.[30] The lattice-resolved image of a single crystalline domain embedded in the amorphous matrix has an interplanar spacing of 0.19 nm, which corresponds to the ($1\bar{1}4$) plane of hexagonal $CoMoP_2$ phase (Figure 2e). This is consistent with the fast-Fourier transform (FFT) image that recorded along with the [$\bar{2}21$] direction (Figure 2f).

*3.3. Electrochemical activation and HER performance*

The electrochemical activation process of the pristine $MoS_2$ for HER was conducted in 1 M KOH solution in a standard three-electrode electrochemical cell. At the current density of 10 mA/cm$^2$, Pt/C and bare Ni foam exhibited HER overpotentials of 26 mV and 200 mV, respectively. (Figure 3a). The HER activity of the pure $MoS_2$ and pristine CMPS are roughly the same, which is 116 and 126 mV, respectively. Furthermore, the linear parts of the polarization curves were fitted by the Tafel equation, yielding Tafel slopes of 30, 109, 67, and 80 mV/dec. for commercial Pt/C, Ni foam, bare $MoS_2$, and pristine CMPS samples, respectively (Figure S3). These results indicate that the pure $MoS_2$ and pristine CMPS have similar HER activities during the initial measurements. Then, a long-term stability test was carried out on the pristine CMPS sample with a constant overpotential of 126 mV without *i*R corrections (Figure S4). Interestingly, instead of remaining unchanged or exhibiting degradation in current densities with increasing measurement time, we found a continuously increasing current density. The polarization curve was recorded after 20 h of measurement, as shown in Figure 3b. The overpotential to deliver a current density is decreased to 110 mV. For a meaningful comparison, a normalized current density increment is defined as $\Delta J/J_0$, where $J_0$ is the current density of pristine CMPS, and $\Delta J$ is the current density increase at -0.25 V.[20] This value was determined to be 115 % after 20 h, and it increased to 221% after another 20 h of further testing. Impressively, the overpotential to produce a current density of 10 mA/cm$^2$ is only ~ 75 mV after activation. This value is better than most reported $MoS_2$ based catalysts (**Table S1**), such as CoS-doped $\beta$-$Co(OH)_2$@amorphous $MoS_{2+x}$ hybrid (143 mV),[31] $MoS_2$/$Ni_3S_2$ heterostructures (110 mV),[32] $MoS_2$/NiCo-layered double hydroxide (78 mV),[33] and metallic-phase $MoS_2$ nanosheets (175 mV).[34] Furthermore, the polarization curve after 40 h plus 5000 cycling test is compared in Figure 3b. The minuscule difference in current density suggests the high long-term stability of our vacancy-rich catalyst in the HER process.

To confirm the increase of active centers, the active surface areas of the catalyst before and after activation were analyzed by their electrochemical double-layer capacitances ($C_{dl}$) (Figure S5-7). As shown in 3a, the $C_{dl}$ increased dramatically from 62.1 to 244 mF cm$^{-2}$ after 20 h activation, and finally to 412 mF cm$^{-2}$ after 40 h, illustrating that the increase in catalytic activity can be attributed to the increased electrochemical surface area of MoS$_2$. As a comparison, pure MoS$_2$, or Ni foam show no activation from the LSV (Figure S8 and S9). Electrochemical impedance spectroscopy (EIS) was performed under the HER conditions (Figure 3d). The Nyquist plots reveal that the charge-transfer resistance shows a significant decrease after activation. This indicates an increase in the number of edge-terminated states or vacancy states that were acting as catalytic centers, which favors electron injection from the Ni foam substrate to the catalyst.

*3.4. Catalyst structure evolution before and after electrochemical activation*

The phase structure, as well as the formation of S vacancies, were examined by X-ray photoelectron spectroscopy (XPS). For the pristine MoS$_2$/CoMoP$_2$ sample, a small S 2*s* peak (226.1 eV) next to the Mo 3 $d_{5/2}$ peak was observed (Figure 4a).[35] Specifically, these doublets with binding energies of 228.6/231.9 eV can be assigned to the Mo 3*d* signal of Mo-P bonding in CoMoP$_2$ (Figure 4a). The peaks located at 229.8 and 233.1 eV with a separation of 3.3 eV can be attributed to the Mo species in the MoS$_2$ phase.[36, 37] Moreover, the Mo 3*d* peaks at higher binding energies (232.5 eV and 235.7 eV) demonstrate slight surface oxidation of the pristine MoS$_2$ surface into MoO$_3$.[38] The S 2*p* peak, on the other hand, can be best fitted with two doublets of S 2$p_{3/2}$ energy states (Figure 4b). The binding energy of 161.5 eV corresponds to the terminal (edge) site of S$^{2-}$ species, which have been identified clearly as active sites for the HER catalytic activity in MoS$_2$.[13] The higher binding energy at 163 eV can be assigned to the bridging disulfide S$_2^{2-}$ ligand or apical S$^{2-}$ ligand.[39, 40] The atomic ratio between S and Mo is determined to be ~ 2 by the above fitting results. Further insight into the pristine MoS$_2$ nanostructure was obtained by examination of the Raman spectrum (Figure 4c). The broad peak between 350 - 420 cm$^{-1}$ belongs to the $E^1_{2g}$ (in a plane motion of Mo and S in opposite directions) and $A_{1g}$ (out of plane motions of S atoms) vibrational modes of hexagonal MoS$_2$ phase, respectively. Of particular note is the dramatic decrease in the relative intensities of $A_{1g}$ to $E^1_{2g}$ in comparison with the bulk phase, suggesting that the as-prepared MoS$_2$ structure is edge-poor.[30, 36] This readily explains the poor initial HER activity as discussed below. Another broad peak centered at 543 cm$^{-1}$ is a feature of the amorphous MoS$_2$ phase, which can be assigned to the $v$-(S-S)$_{terminal}$ and $v$-(S-S)$_{bridging}$ vibrations.[41, 42]

In comparison with the pristine sample, the binding energies for Mo-O and Mo-P bonding remained unchanged (Figure 4d) after activation. The increase in the peak density of Mo-O bonding (232.5 eV) can be explained by surface oxidation, as observed in other systems. Interestingly, the full width at half maximum of the Mo-S peak was larger than that of the pristine sample, and the binding energy shifted to a lower position by 0.11 eV. This means the element Mo has different S coordination and lower valence states should exist after activation. The investigation of S 2$p$ spectra indicates that the intensity was significantly reduced after compared to the Mo 3$d$ peak, while the relative intensity of terminal/edge S was increased by comparison with the basal bridging component (Figure 4e). The atomic ratio between S and Mo was decreased to 1.73 by fitting the spectra, corresponding to an S vacancy concentration of 13%. All these suggest the creation of S vacancies after activation, which serves as active centers for HER.[43] More evidence was provided by the Raman spectra, as shown in Figure 4f. The strong $A_{1g}$ peak reveals that the out of plane vibration is favored, suggests in the increasing of terminal/edge-terminated component.[30, 36] The redshift of $A_{1g}$ (401 cm$^{-1}$) to $E^1_{2g}$ (372 cm$^{-1}$) in comparison with the bulk phase was the result of increases in disorder associated with edge defects.[44, 45] Additionally, a satellite peak was observed beside the $E^1_{2g}$ mode, sitting at smaller wavenumbers of 341 cm$^{-1}$. This further positively confirms that the defects are from S vacancies as expected from previous theoretical predications: the S vacancy could be the localization centers for out of plane vibration and strengthen the $A_{1g}$ mode while acting as scattering centers for in-plane vibration ($E^1_{2g}$ mode) and causing the atoms around the vacancies to have a smaller vibrational amplitude (Figure 4g).[46, 47]

## 4. Conclusions

In conclusion, large concentrations of S vacancies were created in the basal plane of MoS$_2$ when supported on the topological semimetal CoMoP$_2$. Structural characterization revealed that the created vacancies can serve as new catalytic centers for HER and are related to the increase in activity by more than a factor of two, and the efficient electron injection from the electrode to the catalyst. DFT calculations further confirmed these results by revealing a significant decrease in the Gibbs free energy for H$_2$S in the electrochemical process. This synergistic approach could not only be used to design high performance HER catalysts, but also provides a general strategy for creating vacancies for various applications.

## Acknowledgments

The authors gratefully acknowledge Dr. Anil Kumar for the PXRD measurements. We would also like to thank Dr. Qiunan Xu for the helpful discussions. This work was financially supported by the European Research Council (ERC Advanced Grant No. 291472 'Idea Heusler') and ERC Advanced Grant (No. 742068). TOPMAT'. B. Zhang thank the support of the Natural Science Foundation of Shandong Province (CN) with Grants No. ZR2016AB12.

## References

[1] C. Ran, J. Xu, W. Gao, C. Huang, S. Dou, Chem Soc Rev, 47 (2018) 4581-4610.

[2] D.A. Tompsett, S.C. Parker, M.S. Islam, J Am Chem Soc, 136 (2014) 1418-1426.

[3] G. Li, B. Zhang, J. Rao, D. Herranz Gonzalez, G.R. Blake, R.A. de Groot, T.T.M. Palstra, Chem. Mater., 27 (2015) 8220-8229.

[4] D. Ruan, S. Kim, M. Fujitsuka, T. Majima, Appl Catal B-Environ. 288 (2018) 638-646.

[5] J.M. Wu, Y. Chen, L. Pan, P.H. Wang, Y. Cui, D.C. Kong, L. Wang, X.W. Zhang, J.J. Zou, Appl Catal B-Environ, 221 (2018) 187-195.

[6] U. Gupta, C.N.R. Rao, Nano Energy, 41 (2017) 49-65.

[7] H. Wang, Z. Lu, S. Xu, D. Kong, J.J. Cha, G. Zheng, P.C. Hsu, K. Yan, D. Bradshaw, F.B. Prinz, Y. Cui, Proceedings of the National Academy of Sciences of the United States of America, 110 (2013) 19701-19706.

[8] P.S. Maiti, A.K. Ganai, R. Bar-Ziv, A.N. Enyashin, L. Houben, M.B. Sadan, Chem. Mater., 30 (2018) 4489-4492.

[9] S. Chen, X. Chen, G. Wang, L. Liu, Q. He, X.-B. Li, N. Cui, Chem. Mater., 30 (2018) 5404-5411.

[10] Y.Z. Liu, X.Y. Xu, J.Q. Zhang, H.Y. Zhang, W.J. Tian, X.J. Li, M.O. Tade, H.Q. Sun, S.B. Wang, Appl Catal B-Environ. 239 (2018) 334-344.

[11] J. Liu, Y. Liu, D. Xu, Y. Zhu, W. Peng, Y. Li, F. Zhang, X. Fan, Appl Catal B-Environ. 241 (2019) 89-94.

[12] T.F. Jaramillo, K.P. Jorgensen, J. Bonde, J.H. Nielsen, S. Horch, I. Chorkendorff, Science, 317 (2007) 100-102.

[13] G. Li, D. Zhang, Q. Qiao, Y. Yu, D. Peterson, A. Zafar, R. Kumar, S. Curtarolo, F. Hunte, S. Shannon, Y. Zhu, W. Yang, L. Cao, J Am Chem Soc, 138 (2016) 16632-16638.

[14] T.A. Ho, C. Bae, S. Lee, M. Kim, J.M. Montero-Moreno, J.H. Park, H. Shin, Chem. Mater., 29 (2017) 7604-7614.

[15] J. Hong, Z. Hu, M. Probert, K. Li, D. Lv, X. Yang, L. Gu, N. Mao, Q. Feng, L. Xie, J. Zhang, D. Wu, Z. Zhang, C. Jin, W. Ji, X. Zhang, J. Yuan, Z. Zhang, Nat. Commun., 6 (2015) 6293.

[16] D. Liu, Y. Guo, L. Fang, J. Robertson, Appl. Phys. Lett., 103 (2013) 183113.

[17] H. Li, Charlie Tsai, Ai Leen Koh, Lili Cai, Alex W. Contryman, Alex H. Fragapane, Jiheng Zhao, Hyun Soo Han, Hari C. Manoharan, Frank Abild-Pedersen, Jens K. Nørskov, Xiaolin Zheng, Nat. Mater., 15 (2016) 48-53.
[18] Z. Yu, Y. Pan, Y. Shen, Z. Wang, Z.Y. Ong, T. Xu, R. Xin, L. Pan, B. Wang, L. Sun, J. Wang, G. Zhang, Y.W. Zhang, Y. Shi, X. Wang, Nat. Commun., 5 (2014) 5290.
[19] S. Bertolazzi, S. Bonacchi, G. Nan, A. Pershin, D. Beljonne, P. Samori, Adv Mater, 29 (2017) 1606760.
[20] C. Tsai, H. Li, S. Park, J. Park, H.S. Han, J.K. Norskov, X. Zheng, F. Abild-Pedersen, Nat Commun, 8 (2017) 15113.
[21] S. Cristol, J.F. Paul, E. Payen, J. Phys. Chem. B, 104 (2000) 11220-11229.
[22] H. Chen, W. Zhu, D. Xiao, Z. Zhang, Phys Rev Lett, 107 (2011) 056804.
[23] D. Kong, Y. Cui, Nat Chem, 3 (2011) 845-849.
[24] Q.L. He, Y.H. Lai, Y. Lu, K.T. Law, I.K. Sou, Sci Rep, 3 (2013) 2497.
[25] Z.Q. Liu, C.J. Li, W.M. Lü, X.H. Huang, Z. Huang, S.W. Zeng, X.P. Qiu, L.S. Huang, A. Annadi, J.S. Chen, J.M.D. Coey, T. Venkatesan, Ariando, Phy. Rev. X, 3 (2013) 021010.
[26] T. Zhang, Y. Jiang, Z. Song, H. Huang, Y. He, Z. Fang, H. Weng, C. Fang, arXiv:1807.08756 [cond-mat.mtrl-sci], (2018).
[27] Z. Lin, B.R. Carvalho, E. Kahn, R. Lv, R. Rao, H. Terrones, M.A. Pimenta, M. Terrones, 2D Materials, 3 (2016) 022002.
[28] M.G. Sensoy, D. Vinichenko, W. Chen, C.M. Friend, E. Kaxiras, Physical Review B, 95 (2017).
[29] H. Lu, A. Kummel, J. Robertson, APL Mater., 6 (2018) 066104.
[30] M.R. Gao, M.K. Chan, Y. Sun, Nat Commun, 6 (2015) 7493.
[31] T. Yoon, K.S. Kim, Adv. Funct. Mater., 26 (2016) 7386-7393.
[32] J. Zhang, T. Wang, D. Pohl, B. Rellinghaus, R. Dong, S. Liu, X. Zhuang, X. Feng, Angewandte Chemie, 55 (2016) 6702-6707.
[33] J. Hu, C. Zhang, L. Jiang, H. Lin, Y. An, D. Zhou, M.K.H. Leung, S. Yang, Joule, 1 (2017) 383-393.
[34] X. Geng, W. Sun, W. Wu, B. Chen, A. Al-Hilo, M. Benamara, H. Zhu, F. Watanabe, J. Cui, T.P. Chen, Nat Commun, 7 (2016) 10672.
[35] J. Kibsgaard, T.F. Jaramillo, Angewandte Chemie, 53 (2014) 14433-14437.
[36] M.A.R. Anjum, H.Y. Jeong, M.H. Lee, H.S. Shin, J.S. Lee, Adv Mater, 30 (2018) e1707105.
[37] J. Wang, J. Luo, D. Liu, S. Chen, T. Peng, Appl Catal B-Environ. 241 (2019) 130-140.
[38] G. Li, Y. Sun, J. Rao, J. Wu, A. Kumar, Q. Xu, C. Fu, E. Liu, R.G. Blake, P. Werner, B. Shao, K. Liu, S. Parkin, X. Liu, M. Fahlman, S.-C. Liou, G. Auffermann, J. Zhang, C. Felser, X. Feng, Adv. Energy. Mater., 1801258 (2018) DOI: 10.1002/aenm.201801258.
[39] D. Dinda, M.E. Ahmed, S. Mandal, B. Mondal, S.K. Saha, J Mater. Chem. A, 4 (2016) 15486-15493.
[40] Y. Li, K. Yin, L.L. Wang, X.L. Lu, Y.Q. Zhang, Y.T. Liu, D.F. Yan, Y.Z. Song, S.L. Luo, Appl Catal B-Environ. 239 (2018) 537-544.
[41] Y. Deng, L.R.L. Ting, P.H.L. Neo, Y.-J. Zhang, A.A. Peterson, B.S. Yeo, ACS Catalysis, 6 (2016) 7790-7798.

[42] P.D. Tran, T.V. Tran, M. Orio, S. Torelli, Q.D. Truong, K. Nayuki, Y. Sasaki, S.Y. Chiam, R. Yi, I. Honma, J. Barber, V. Artero, Nat Mater, 15 (2016) 640-646.

[43] A.Y. Lu, X. Yang, C.C. Tseng, S. Min, S.H. Lin, C.L. Hsu, H. Li, H. Idriss, J.L. Kuo, K.W. Huang, L.J. Li, Small, 12 (2016) 5530-5537.

[44] J. Benson, M. Li, S. Wang, P. Wang, P. Papakonstantinou, ACS Appl Mater Interfaces, 7 (2015) 14113-14122.

[45] D. Gopalakrishnan, D. Damien, M.M. Shaijumon, ACS Nano, 8 (2014) 5297-5303.

[46] S. Bae, N. Sugiyama, T. Matsuo, H. Raebiger, K.-i. Shudo, K. Ohno, Phys. Rev. Appl., 7 (2017).

[47] K. Wu, Z. Li, J. Tang, X. Lv, H. Wang, R. Luo, P. Liu, L. Qian, S. Zhang, S. Yuan, Nano Research, 11 (2018) 4123-4132.

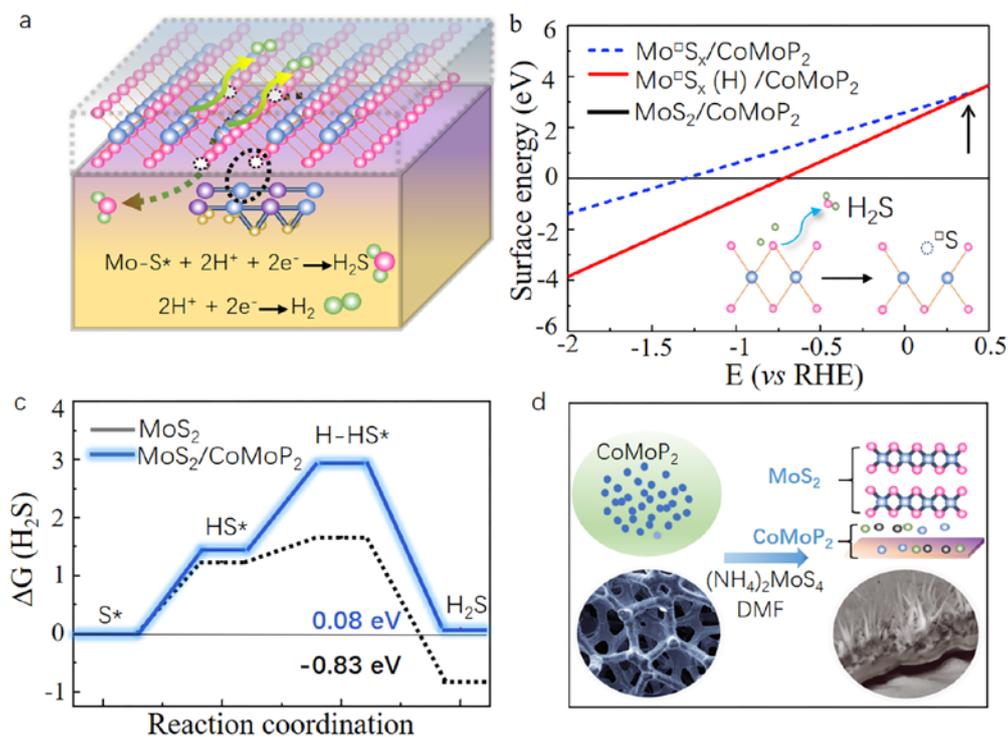

**Figure 1** (a). Illustration of the creation of S vacancies in the basal plane of $MoS_2$. S is removed with a two-step process in the form of $H_2S$. The created S vacancies can serve as active sites for HER. (b). The calculated surface free energy of pristine $MoS_2/CoMoP_2$ (Black line), $MoS_2/CoMoP_2$ with S vacancies ($Mo^\Box S_x/CoMoP_2$), and with a hydrogen adsorbed on the S vacancy ($M^\Box S_x (H) / CoMoP_2$). (c). Comparison of the Gibbs free energy for the formation of $H_2S$ between $MoS_2/CoMoP_2$ and pure phase of $MoS_2$. (d). The syntactic strategy for the semimetal supported $MoS_2$.

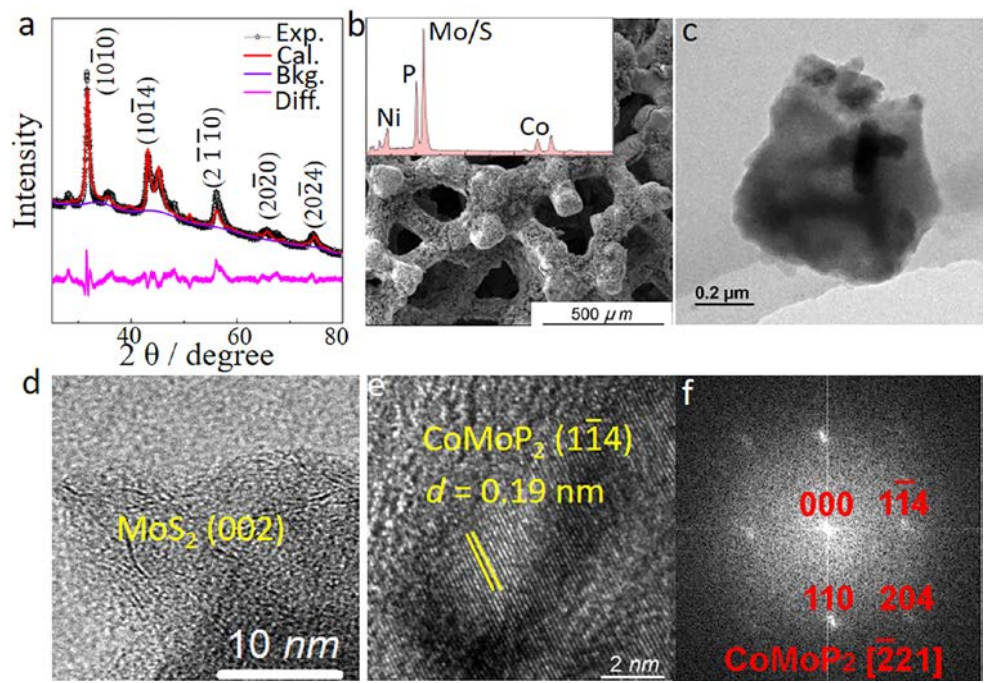

**Figure 2.** (a). The XRD pattern of the synthesized MoS$_2$/CoMoP$_2$ sample. (b). SEM and (c). TEM image of the MoS$_2$/CoMoP$_2$ particle. (d). HRTEM shows the low crystalline MoS$_2$ phase outside, and (e). Semimetal CoMoP$_2$ phase inside, and (f). the corresponding fast Fourier transform (FFT).

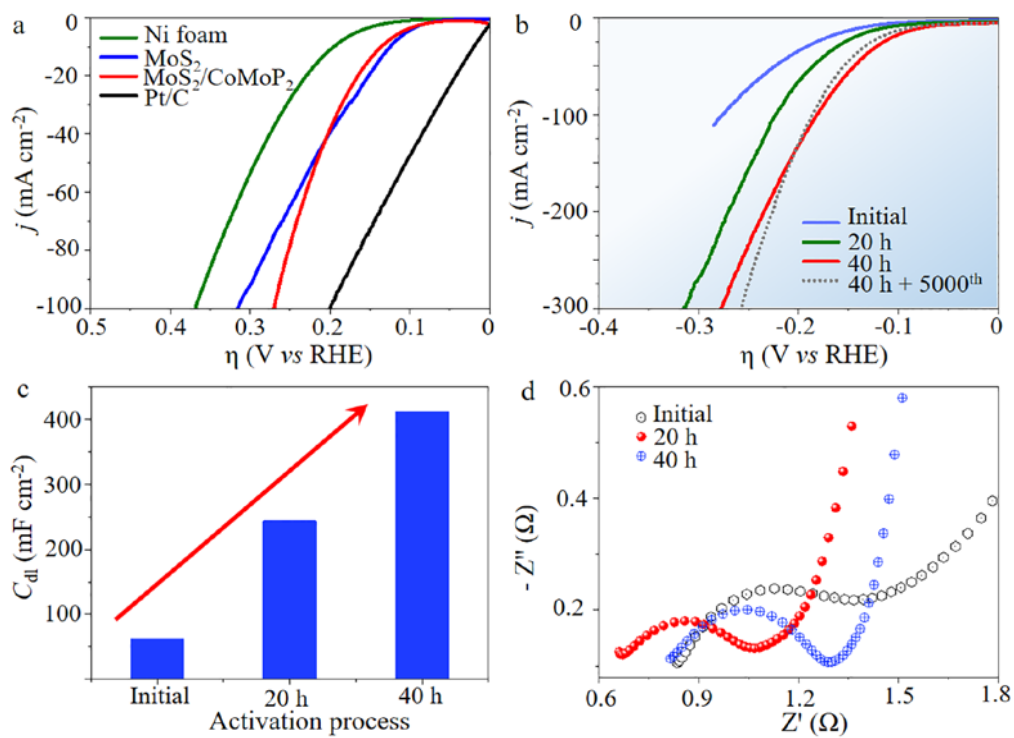

**Figure 3**. (a). Polarization curves of Ni foam (NF), pure $MoS_2$ phase, commercial Pt/C catalyst, and pristine $MoS_2/CoMoP_2$. (b). Comparison of polarization curves for the $MoS_2/CoMoP_2$ catalyst in the initial test, and after activation for 20 h, 40 h, and 40 h plus 5000 cycles. (c). Plot showing the increase of double-layer capacitance ($C_{dl}$) for the pristine and vacancy-rich $MoS_2/CoMoP_2$ sample for the initial test, activated for 20h, and 40 h, respectively. (d). Impedance measurement of the pristine $MoS_2/CoMoP_2$ and activated for 20 and 40 h, respectively.

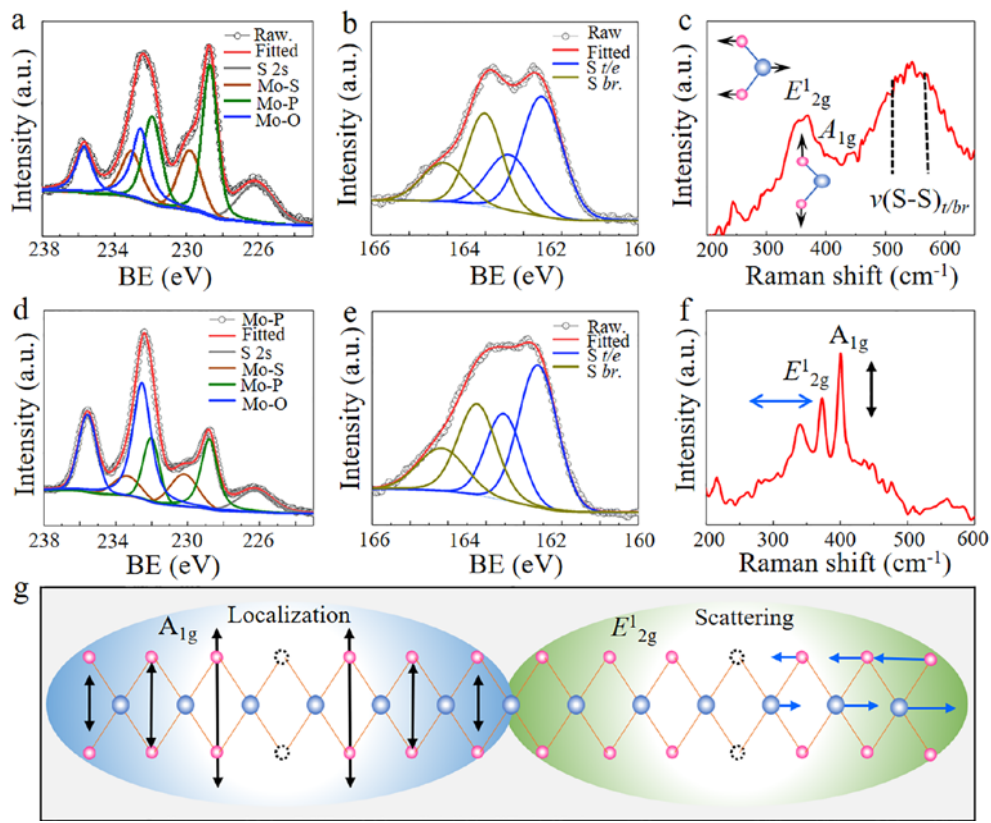

**Figure 4.** (a). XPS spectra of Mo 3d peak, and (b). S 2p peak, for the pristine $MoS_2/CoMoP_2$ sample. (c). Raman spectrum of the pristine $MoS_2/CoMoP_2$. XPS spectra of (d). Mo 3d peak, and (e). S 2p peak, for the activated $MoS_2/CoMoP_2$ sample. (f). Raman spectrum of the $MoS_2/CoMoP_2$ after activation. (g). An illustration of the function of S vacancies as localization and scattering centers for Raman vibration.